\documentclass[sigconf,screen=true,anonymous=false,bookmarks=false,nonacm]{acmart}
\iftrue
\usepackage{titlesec}
\titlespacing\section{2pt}{5pt plus 1pt minus 1pt}{0pt plus 1pt minus 1pt}
\titlespacing\subsection{2pt}{5pt plus 1pt minus 1pt}{0pt plus 1pt minus 1pt}
\titlespacing\subsubsection{2pt}{5pt plus 1pt minus 1pt}{2pt plus 1pt minus 1pt}
\usepackage{enumitem}
\setlist{leftmargin=5.08mm}
\fi

\iftrue
\fi

\usepackage{lipsum}
\usepackage{graphicx}
\usepackage{amsmath}
\usepackage{footnote}
\usepackage{mathtools}
\usepackage{mathrsfs}     
\usepackage{comment}
\usepackage[subrefformat=parens,labelformat=parens]{subfig}
\usepackage{bm,bbm}
\usepackage{multirow}
\usepackage{threeparttable,booktabs}
\usepackage{blkarray}
\usepackage{tikz}
\usetikzlibrary{positioning,calc,fit,decorations.pathmorphing,shapes.geometric,shapes.gates.logic.US,calc}
\usepackage{tikz-3dplot}
\usepackage{balance}
\usepackage{courier}                                       
\usepackage{cleveref}                                      
\usepackage[mathcal]{eucal}
\usepackage[]{algpseudocode}                               
\algrenewcommand\textproc{\texttt}
\makeatletter\let\float@addtolists\relax\makeatother
\usepackage{algorithm}
\usepackage{filecontents}                                  
\usepackage{pgfplots}
\usepackage{pgfplotstable}
\usepackage{pgf-pie}
\usepgfplotslibrary{groupplots}
\pgfplotsset{compat=newest}
\usepackage[figuresright]{rotating}
\usepackage{xcolor,colortbl}                               

\theoremstyle{plain}

\theoremstyle{definition}

\algrenewcommand\textproc{\texttt}

\usepackage[skip=1pt]{caption}            
\definecolor{CUHKorange}{RGB}{244,106,18} 
\definecolor{CUHKblue}{RGB}{0,111,190}    
\definecolor{CUHKgreen}{RGB}{0,127,128}   
\definecolor{CUHKred}{RGB}{228,46,36}     
\definecolor{CUHKyellow}{RGB}{198,148,34} 
\definecolor{CUHKdark}{RGB}{114,44,114}   
\definecolor{CUHKmiddle}{RGB}{144,44,144} 

\usepackage{tcolorbox}
\tcbuselibrary{skins,breakable}
    {\endtcolorbox}

\graphicspath{{./figs/}{../}}
\usepackage{tikz}
\usepackage{pgfplots}

\usepackage{svg}
\usepackage{expl3}
\usepackage{arydshln}
\usepackage{diagbox}
\usepackage{tikz}
\usepackage[commentColor=blue,noEnd=false]{algpseudocodex}
\tikzset{algpxIndentLine/.style={draw=black}}
\usepackage{bm}
\usetikzlibrary{arrows.meta}
\makeatletter

\definecolor{mygold}{HTML}{b05f8c}
\definecolor{mypurple}{HTML}{f7d35c}
\definecolor{mypink}{RGB}{255,171,164}
\definecolor{C0}{HTML}{8dd3c7}
\definecolor{C1}{HTML}{ffffb3}
\definecolor{C2}{HTML}{bebada}
\definecolor{C3}{HTML}{fb8072}
\definecolor{C4}{HTML}{80b1d3}
\definecolor{C5}{HTML}{fdb462}
\definecolor{C6}{HTML}{b3de69}

\definecolor{C0}{HTML}{b05f8c}
\definecolor{C1}{HTML}{f58297}
\definecolor{C2}{HTML}{f7d35c}

\begin{document}
\date{}

\title{
FluxEDA: A Unified Execution Infrastructure for Stateful Agentic EDA
}

\author{
    Zhengrui Chen, Zixuan Song, Yu Li, Qi Sun, Cheng Zhuo \\
    Zhejiang University, Hangzhou, China
}

\begin{abstract}

Large language models and autonomous agents are increasingly explored for EDA automation, but many existing integrations still rely on script-level or request-level interactions, which makes it difficult to preserve tool state and support iterative optimization in real production-oriented environments. In this work, we present FluxEDA, a unified and stateful infrastructure substrate for agentic EDA. 
FluxEDA introduces a managed gateway-based execution interface with structured request and response handling. It also maintains persistent backend instances. Together, these features allow upper-layer agents and programmable clients to interact with heterogeneous EDA tools through preserved runtime state, rather than through isolated shell invocations. We evaluate the framework using two representative commercial backend case studies: automated post-route timing ECO and standard-cell sub-library optimization.
The results show that FluxEDA can support multi-step analysis and optimization over real tool contexts, including state reuse, rollback, and coordinated iterative execution. These findings suggest that a stateful and governed infrastructure layer is a practical foundation for agent-assisted EDA automation.

\end{abstract}
\maketitle
\pagestyle{plain}
\pagenumbering{gobble}
\section{Introduction}

With the continuous scaling down of IC feature sizes and the exponential growth in transistor counts, modern chip design has become overwhelmingly complex. This unprecedented scale necessitates a heavy reliance on Electronic Design Automation (EDA) tools across the entire chip design flow~\cite{huang2021edasurvey}. In industrial practice, dedicated EDA engineers orchestrate these complex workflows by relying on custom scripts to manually integrate disparate engines (e.g., synthesis, placement, routing, and timing) into a cohesive physical design flow. However, this flow integration work is highly repetitive, labor-intensive, and prone to human error~\cite{chen2001scripting}. EDA teams often expend significant engineering resources to maintain baseline script functionality and handle environment configurations, rather than focusing on exploring optimal design spaces or developing advanced heuristics~\cite{chen2024dawn}.

Recent advances in large language models (LLMs) and autonomous agents present new opportunities to alleviate this burden and accelerate research and development in the IC design flow. Agentic AI systems possess the theoretical capability to comprehend high-level specifications, autonomously execute multi-step toolchains, and iteratively refine design parameters~\cite{zang2026dawnagenticedasurvey}. Early frameworks, such as ChatEDA \cite{wu2024chateda}, attempt to harness this by generating tool-specific scripts. However, modern EDA tools, encompassing both open-source academic frameworks and industrial-grade commercial engines, are highly isolated. They utilize disparate interfaces, proprietary shell environments (\textit{e.g.}, Tcl~\cite{openroad2019,li2023ieda,chen2025zlibboost}, Python~\cite{empyrean_2024}, Bash~\cite{iverilog,verilator,BraytonMishchenko2010abc}), and disconnected data formats. Directly interfacing AI agents with this heterogeneous ecosystem creates a prohibitive $\mathbf{N \times M}$ integration bottleneck.

To resolve this, standardizations, such as Model Context Protocol (MCP)~\cite{mcp}, have emerged. Recent works (\textit{e.g.}, MCP4EDA~\cite{wang2025mcp4eda}, AutoEDA~\cite{lu2025autoeda}) employ MCP or related microservice concepts to reduce integration complexity across heterogeneous EDA tools. Nevertheless, in practice, most existing frameworks still couple the agent to EDA engines through script-level or request-level batch interactions, where each step is mediated by generated commands and post-hoc log parsing. Such an interaction model can support workflow orchestration, but it does not fully provide the persistent, structured, and governable tool state needed for backend physical design~\cite{kahng2024solvers}. Consequently, agents remain limited in their ability to perform fine-grained, continuous optimization over real tool contexts, especially in production-oriented commercial environments.

Furthermore, blindly executing these LLM-generated scripts without fine-grained runtime governance poses severe reliability risks in industrial settings. As highlighted by Andrew Kahng et al.~\cite{ghose2026agentic}, current agentic EDA frameworks lack strict semantic boundaries, granting LLMs unrestricted access to the underlying execution environments~\cite{zang2026dawnagenticedasurvey}. This unconstrained action space leaves the physical design process vulnerable to hallucinated commands, silent state corruption, and unpredictable tool crashes, fundamentally limiting their viability in production-critical workflows.

\begin{figure}[t]
    \centering
    \includegraphics[width=1.0\linewidth]{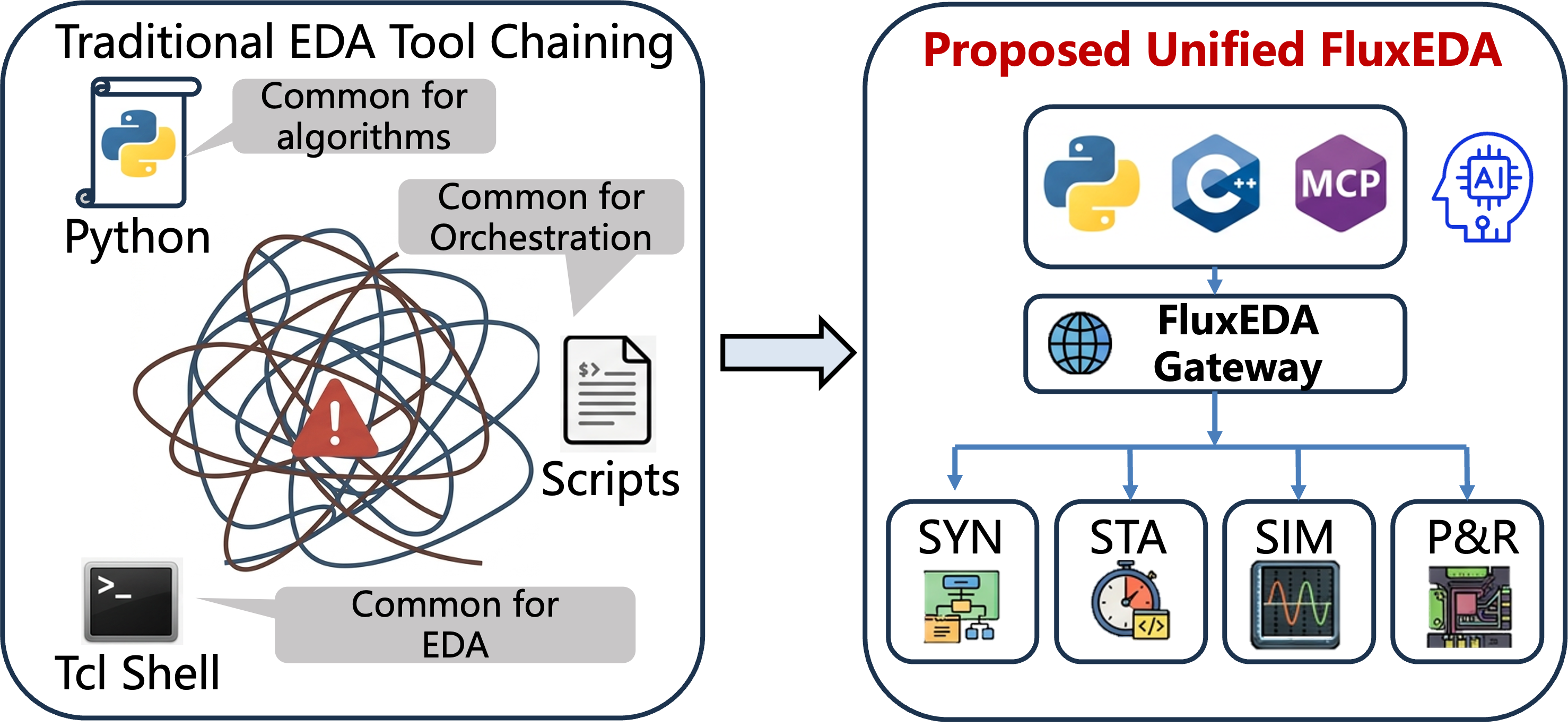}
    \caption{Comparison between traditional EDA workflows and the proposed unified FluxEDA, highlighting the shift from fragmented scripting and tool-specific orchestration to a cohesive, standardized framework for EDA tool integration.}
    \label{fig:eda_comparison}
\end{figure}

\begin{figure*}[h!]
    \centering
    \includegraphics[width=0.7\linewidth]{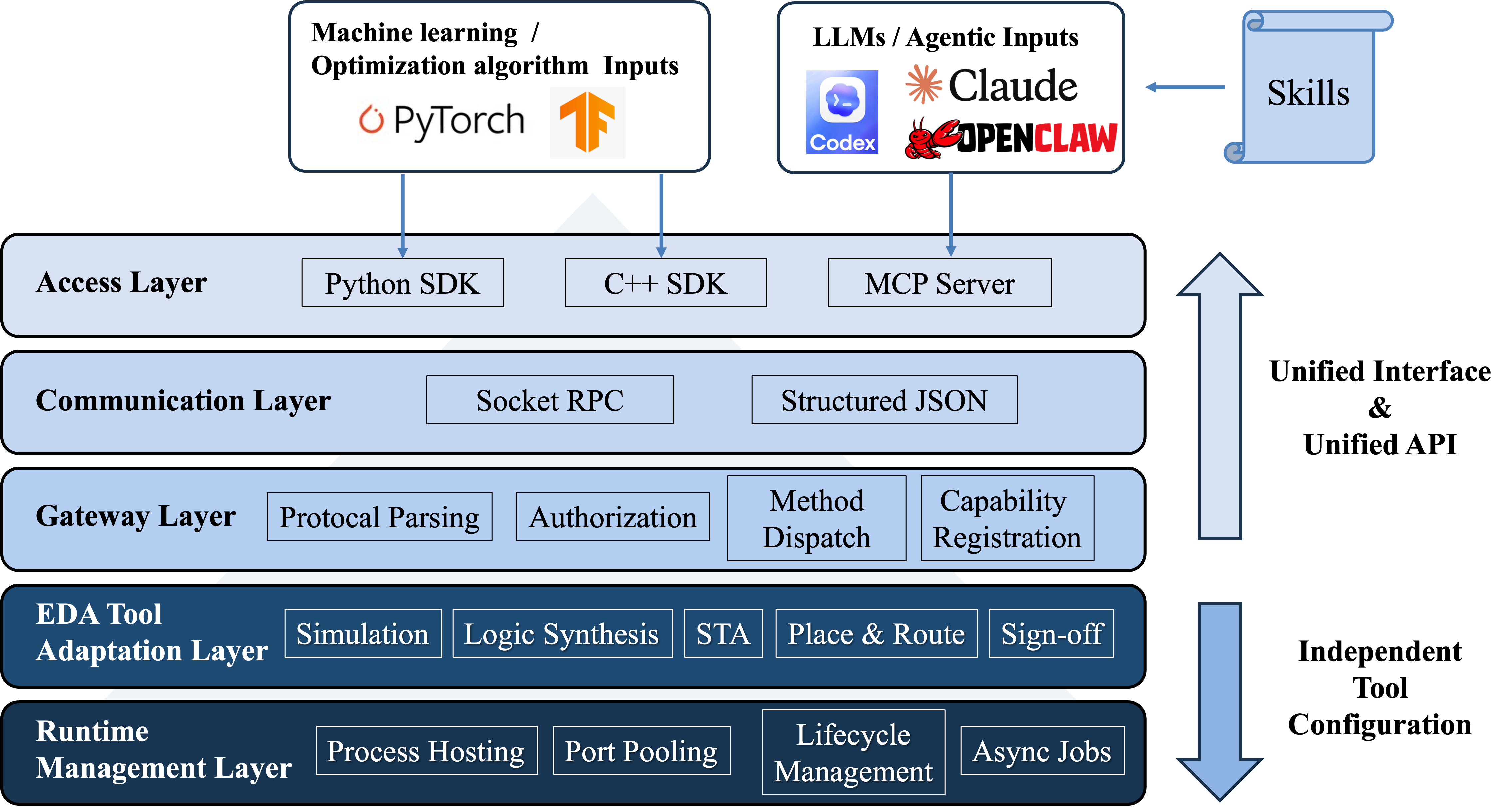}
    \caption{Overall software architecture of FluxEDA. The system organizes heterogeneous EDA environments into a five-layer stack that provides unified access, controlled capability exposure, and persistent tool execution for agentic and programmable workflows.}
    \label{fig:arch}
\end{figure*}

To address these limitations, we present ~\textbf{FluxEDA}, a unified infrastructure substrate for reliable and stateful agentic EDA. Rather than exposing heterogeneous EDA tools directly through fragmented scripts or ad hoc shell interfaces, FluxEDA inserts a stable capability layer between upper-layer clients and the underlying tool environments, as shown in Figure~\ref{fig:eda_comparison}. At its core, the FluxEDA Gateway encapsulates tool-native functionalities via a structured Socket-RPC interface and provides a managed execution substrate for tool interaction. The gateway is primarily accessed through an MCP Server for standardized agent-facing integration, while Python and C++ SDKs are also supported for programmatic use. Through this design, heterogeneous engines spanning logic synthesis, physical design, and verification are absorbed into a unified API space within the digital implementation flow. This allows agents to operate over persistent tool sessions rather than discrete script boundaries, while also enabling conventional software clients to interact with the same managed runtime. Furthermore, FluxEDA incorporates robust platform-level runtime management, including instance routing, heartbeat-based session maintenance, timeout control, and asynchronous long-task execution, thereby enabling stable long-horizon interaction within production-oriented EDA contexts.

While the MCP server and gateway establish a persistent tool-access surface, raw API exposure alone is insufficient for complex digital implementation. To bridge the gap between low-level tool invocation and high-level design intent, FluxEDA introduces domain-specific Skills~\cite{anthropic_claude_skills_docs_2026} to encode explicit workflow knowledge. Rather than relying entirely on the agent's implicit reasoning to navigate intricate EDA processes, Skills provide structured procedural guidance. This encompasses task decomposition, input validation, execution ordering, session-context preservation, and reusable operational templates. Consequently, this architecture enforces a strict separation of concerns: the MCP layer governs how tool capabilities are safely exposed and managed, whereas the Skills layer dictates when and in what sequence these capabilities should be composed to accomplish concrete flow tasks.

To validate our proposed architecture, we instantiate FluxEDA across industrial commercial EDA environments. We demonstrate the framework's effectiveness in two highly iterative, representative backend scenarios: automated post-route timing ECO closure and Pareto-driven sub-library optimization. The main contributions are summarized as follows:
\begin{itemize}
 \item We present \textbf{FluxEDA}, a unified and stateful infrastructure substrate for agentic EDA that standardizes heterogeneous tool access through a gateway-based architecture with persistent sessions, explicit capability registration, and platform-level runtime management.
 \item We introduce a layered interaction model that separates low-level tool access from high-level workflow knowledge: MCP provides a standardized and governable tool-access surface, while domain-specific Skills encode reusable task decomposition, input constraints, execution ordering, and session-aware workflow guidance.
 \item We instantiate and evaluate FluxEDA on two representative case studies using commercial EDA environments. This demonstrates that the framework natively supports continuous analyze–execute–refine loops over persistent, in-memory tool contexts, overcoming the limitations of one-shot, black-box script execution prevalent in practical EDA workflows.
\end{itemize}

The remainder of this paper is organized as follows. \Cref{sec:arch} details our framework. \Cref{sec:exp} is for experiments. Finally, \Cref{sec:conclu} concludes this paper.

\section{Software Architecture}
\label{sec:arch}

\subsection{Overview}
\begin{figure*}[h]
    \centering
    \includegraphics[width=0.9\linewidth]{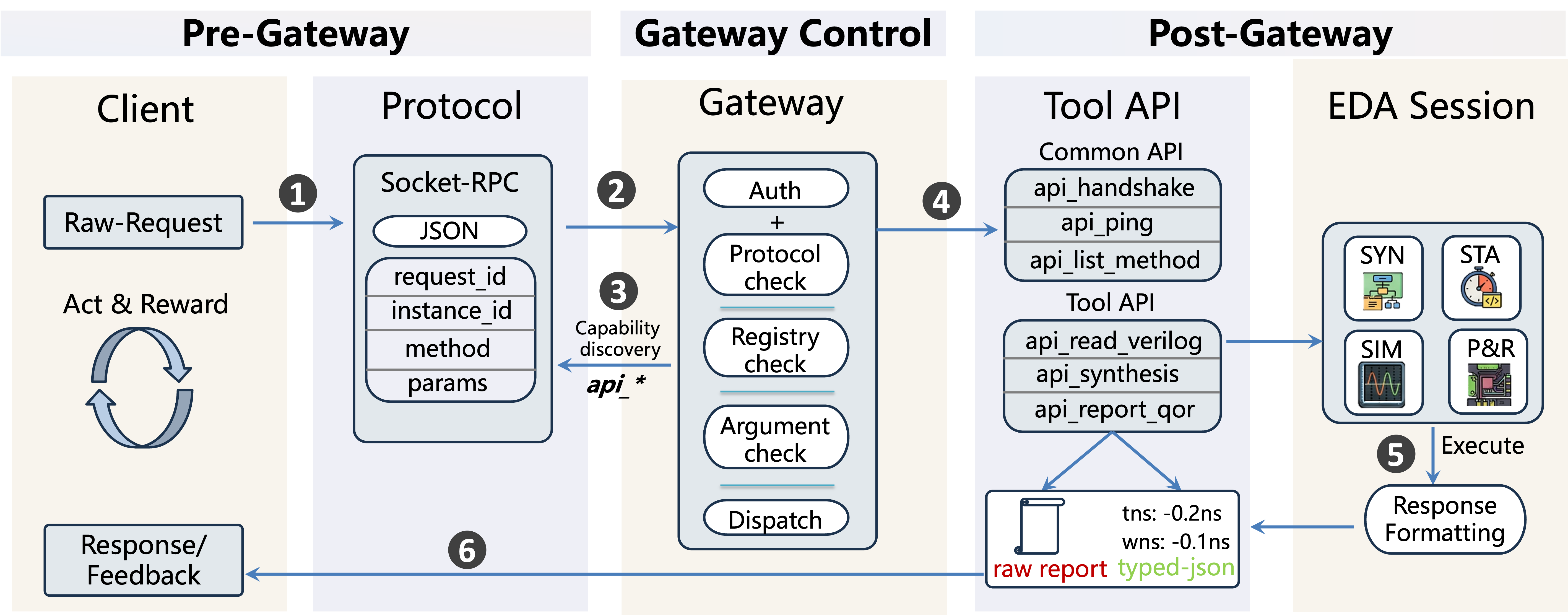}
    \caption{Execution signal flow in FluxEDA. Agent requests are issued through task logic, forwarded by the RPC client and session manager to the Tcl gateway, resolved through registered \texttt{api\_*} methods, and executed on managed EDA runtimes. Capability discovery, method metadata, typed results, artifacts, and execution status traverse the same structured path.}
    \label{fig:dataflow}
\end{figure*}

Figure~\ref{fig:arch} shows the overall architecture of FluxEDA. The system is organized into five layers: Access, Communication, Gateway, EDA Tool Adaptation, and Runtime Management. Rather than directly exposing tool-specific shells to the user or agent, FluxEDA inserts a common execution stack between frontend clients and backend EDA environments. This organization provides a consistent invocation path across different interfaces while keeping tool-specific execution and session management inside the infrastructure layer.

From an EDA perspective, the main purpose of this architecture is to support multi-step and cross-tool interaction under a persistent execution model. Many practical tasks are not single commands, but sequences of dependent analysis, reporting, and refinement steps that must operate on the same loaded design and runtime state. FluxEDA therefore combines unified access, controlled method dispatch, and managed tool sessions into a single software substrate for programmable and agent-assisted EDA automation.

\subsection{Execution Signal Flow}

Fig.~\ref{fig:dataflow} illustrates the end-to-end request path in FluxEDA as a numbered signal flow spanning the pre-gateway, gateway-control, and post-gateway stages. At step~(1), an incoming request is packaged into a structured socket-based RPC message carrying the target method, arguments, and execution metadata, and is then forwarded into the common communication and gateway path.

After crossing the RPC boundary, the request enters the gateway control stage. At step~(2), FluxEDA applies the control checks required before execution, including protocol validation, registry lookup, argument checking, and capability-level authorization. Rather than exposing arbitrary shell procedures directly, the gateway resolves requests only against explicitly registered \texttt{api\_*} methods, which defines a clear execution boundary for backend access. Gateway also supports capability discovery at step~(3), where clients can query connectivity, available methods, and associated interface metadata through common introspection APIs.

Validated executable requests are dispatched at step~(4) to the corresponding common or tool-specific APIs. At step~(5), the target backend instance executes the requested operation and the result is normalized along the post-gateway path. Finally, at step~(6), FluxEDA returns a structured response to the caller, including status information and, depending on the selected return mode, typed data, text-like outputs, or raw artifacts. By using a unified communication contract for invocation, introspection, and result delivery, FluxEDA keeps upper-layer automation aligned with the executable interface and reduces dependence on tool-specific shell conventions.

\begin{figure}[h]
    \centering
    \includegraphics[width=0.7\linewidth]{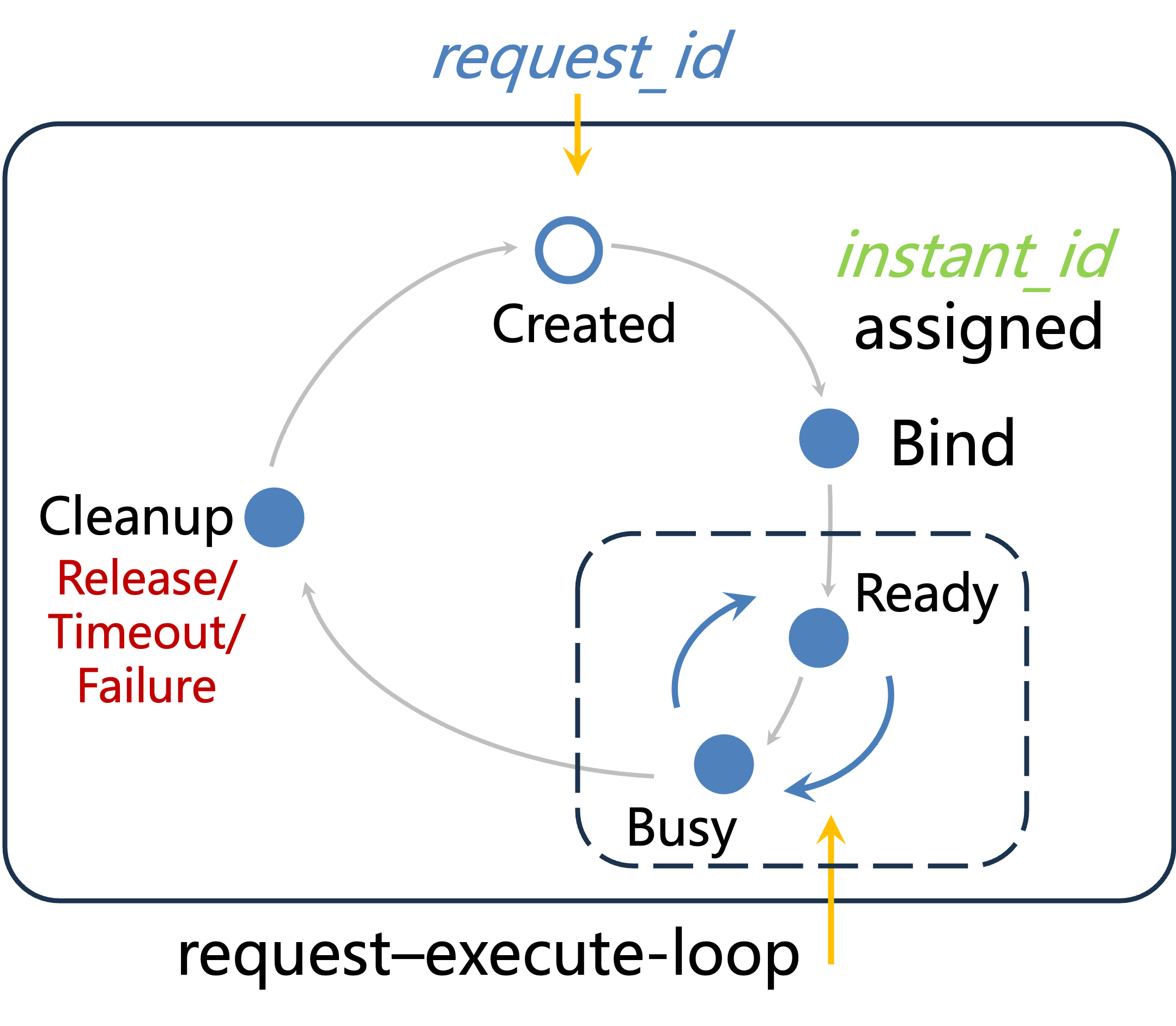}
    \caption{Lifecycle of a managed backend instance in FluxEDA. After a tool process is created and bound to an \texttt{instance\_id}, multiple requests can be routed to the same live instance before it is eventually reclaimed by runtime management.}
    \label{fig:lifecycle}
\end{figure}

\subsection{Progressive Capability Exposure and Skills}

A practical challenge in agent-facing EDA infrastructure is how to expose tool capabilities without overwhelming the upper-layer client. In practice, a flat MCP interface that publishes all callable methods and their descriptions upfront not only becomes increasingly heavy as backend capabilities grow, but also consumes valuable context budget for LLM-based agents~\cite{hasan2025mcpmodel}. To address this issue, FluxEDA adopts a progressive capability exposure mechanism: the gateway first provides only a compact discovery surface, e.g., through introspection APIs such as \texttt{api\_ping} and \texttt{api\_list\_method}, while detailed invocation requirements are queried only when needed. This design keeps the protocol surface compact and extensible as backend tools evolve. Meanwhile, workflow knowledge is handled separately through domain-specific Skills, which guide capability selection and invocation ordering over the discovered method space without overloading the protocol layer itself.

\subsection{Tool Adaptation and Lifecycle Management}

Below the gateway, FluxEDA uses tool adapters to expose native tool capabilities through a consistent callable interface. The purpose of this layer is not to erase tool-specific behavior, since timing analysis, synthesis, simulation, and implementation tools naturally differ in command semantics, state models, and report formats. Instead, these differences are confined behind registered \texttt{api\_*} methods, allowing upper layers to invoke backend operations without depending directly on tool-specific shell conventions.

This interface abstraction is coupled with lifecycle-managed backend instances. Rather than treating each request as an isolated shell invocation, FluxEDA reuses managed instances that preserve the loaded design, execution context, and intermediate state across multiple calls. This execution model is essential for practical EDA automation, where a task often spans dependent steps such as design loading, report inspection, timing analysis, optimization, and re-analysis under updated state.

Figure~\ref{fig:lifecycle} illustrates this mechanism. A backend tool process is first brought under runtime management and bound to an \texttt{instance\_id}, which serves as the routing handle for subsequent calls. Once established, the same instance can be reused by a sequence of requests while preserving the loaded design and internal tool context across interactions. In contrast, each RPC carries a distinct \texttt{request\_id} for per-call tracking, response matching, and error reporting. This separation allows FluxEDA to maintain both instance continuity and request-level observability within a single invocation framework.

The runtime management layer sustains this lifecycle through process startup, instance binding and routing, port allocation, readiness checks, liveness monitoring, support for long-running operations, and cleanup on explicit release, timeout, or failure. As a result, higher-level automation can operate on stable backend instances without repeatedly reconstructing the tool environment. This persistent-instance execution model distinguishes FluxEDA from a simple RPC wrapper: FluxEDA standardizes not only how backend tools are invoked, but also how their runtime state is created, preserved, reused, and reclaimed for iterative and cross-tool EDA tasks.

\begin{table*}[htbp]
\centering
\caption{Automated post-route timing ECO results. Iteration~2 is rejected because more aggressive setup sizing does not improve WNS and slightly degrades TNS and the number of setup-violating paths. Iteration~3 resumes from the preserved Iteration~1 state and performs targeted hold repair.}
\label{tab:eco_timing_case}
\renewcommand{\arraystretch}{1.15}
\setlength{\tabcolsep}{8pt}
\begin{tabular*}{\textwidth}{@{\extracolsep{\fill}}lrrrrrr}
\toprule
State & Setup WNS (ns) & Setup TNS (ns) & Setup Violating Paths & Hold WHS (ns) & Hold THS (ns) & Hold Violations \\
\midrule
Baseline & -0.81 & -37.37 & 140 & -0.39 & -1.33 & 4 \\
Iteration 1 & -0.76 & -34.78 & 130 & -0.38 & -1.33 & 4 \\
Iteration 2 & -0.76 & -34.84 & 133 & -0.39 & -1.33 & 4 \\
Iteration 3 & \textbf{-0.76} & \textbf{-34.78} & \textbf{130} & \textbf{+0.01} & \textbf{0.00} & \textbf{0} \\
\bottomrule
\end{tabular*}
\end{table*}

\begin{table*}[htbp]
\centering
\caption{Key milestones in Pareto-driven standard-cell sub-library selection. The last two columns report the number of exact cell references and cell families used in each synthesized implementation.}
\label{tab:lib_subset_case}
\renewcommand{\arraystretch}{1.15}
\setlength{\tabcolsep}{7pt}
\begin{tabular}{lrrrrrr}
\toprule
State & Area ($\mu\mathrm{m}^2$) & WNS (ns) & TNS (ns) & Viol. Paths & Used Cell Refs. ($\downarrow$ vs. Base) & Used Cell Families. ($\downarrow$ vs. Base) \\
\midrule
Baseline & 14651.55 & -0.246 & -20.996 & 122 & 149 (0.0\%)  & 51 (0.0\%)  \\
A-Run9   & 12530.89 & -1.114 & -186.934 & 237 & 22 (85.2\%)  & 20 (60.8\%) \\
T-Run04  & 12748.21 & -0.791 & -97.780  & 228 & 23 (84.6\%)  & 20 (60.8\%) \\
T-Run10  & 14580.79 & -0.385 & -41.803  & 181 & 30 (79.9\%)  & 20 (60.8\%) \\
\textbf{T-Run11}  & \textbf{13659.95} & \textbf{-0.448} & \textbf{-40.306} & \textbf{157} & \textbf{36 (75.8\%)} & \textbf{23 (54.9\%)} \\
\bottomrule
\end{tabular}
\end{table*}

\section{Case Studies}
\label{sec:exp}

We evaluate FluxEDA through two representative backend case studies within commercial EDA environments. The system utilizes GPT-5.4~\cite{openai2026gpt54systemcard} as the primary large language model, while the agent orchestration is implemented using the OpenAI Codex framework~\cite{openai2025codex}. Our goal is to demonstrate that the framework natively supports multi-step optimization over persistent tool sessions. Specifically, we explore automated post-route timing ECO and standard-cell sub-library optimization to highlight the system's proficiency in both continuous refinement and broad automated search.

\subsection{Automated Post-Route Timing ECO Closure}

Our first case study evaluates FluxEDA on automated post-route timing ECO. This task requires continuous diagnosis, repair, and validation over a persistent in-memory design context, rather than a sequence of disconnected batch invocations. The agent first loads the routed netlist, extracted parasitics, and timing constraints, and then establishes a baseline timing view by generating global timing summaries and representative critical-path reports. On the evaluated design, the baseline exhibits setup WNS/TNS of $-0.81/-37.37$ ns over 140 violating paths, together with hold WHS/THS of $-0.39/-1.33$ ns over 4 violating paths. These reports indicate that setup is the dominant bottleneck and is primarily amenable to cell resizing, whereas the remaining hold issues are small in number and can be deferred to a later cleanup stage.

Table~\ref{tab:eco_timing_case} summarizes the optimization trajectory. In Iteration~1, default setup sizing delivers the main setup gain, improving setup WNS to $-0.76$ ns and setup TNS to $-34.78$ ns, while reducing setup-violating paths from 140 to 130. Iteration~2 explores a more aggressive setup-sizing configuration, but yields no further WNS improvement and slightly worsens both TNS and the number of violating paths, indicating diminishing returns. FluxEDA therefore rolls back to the preserved Iteration~1 state and applies targeted hold repair in Iteration~3. This final step eliminates all remaining hold violations, improving hold WHS from $-0.39$ ns to $+0.01$ ns and hold THS from $-1.33$ ns to $0.00$ ns, without sacrificing the earlier setup gains. 

Overall, this case study shows that FluxEDA supports a persistent analyze--execute--refine loop with branch exploration and rollback, allowing the agent to resume from the best intermediate state instead of reconstructing the execution context after each trial.

\begin{figure}[t]
\centering
\includegraphics[width=\columnwidth]{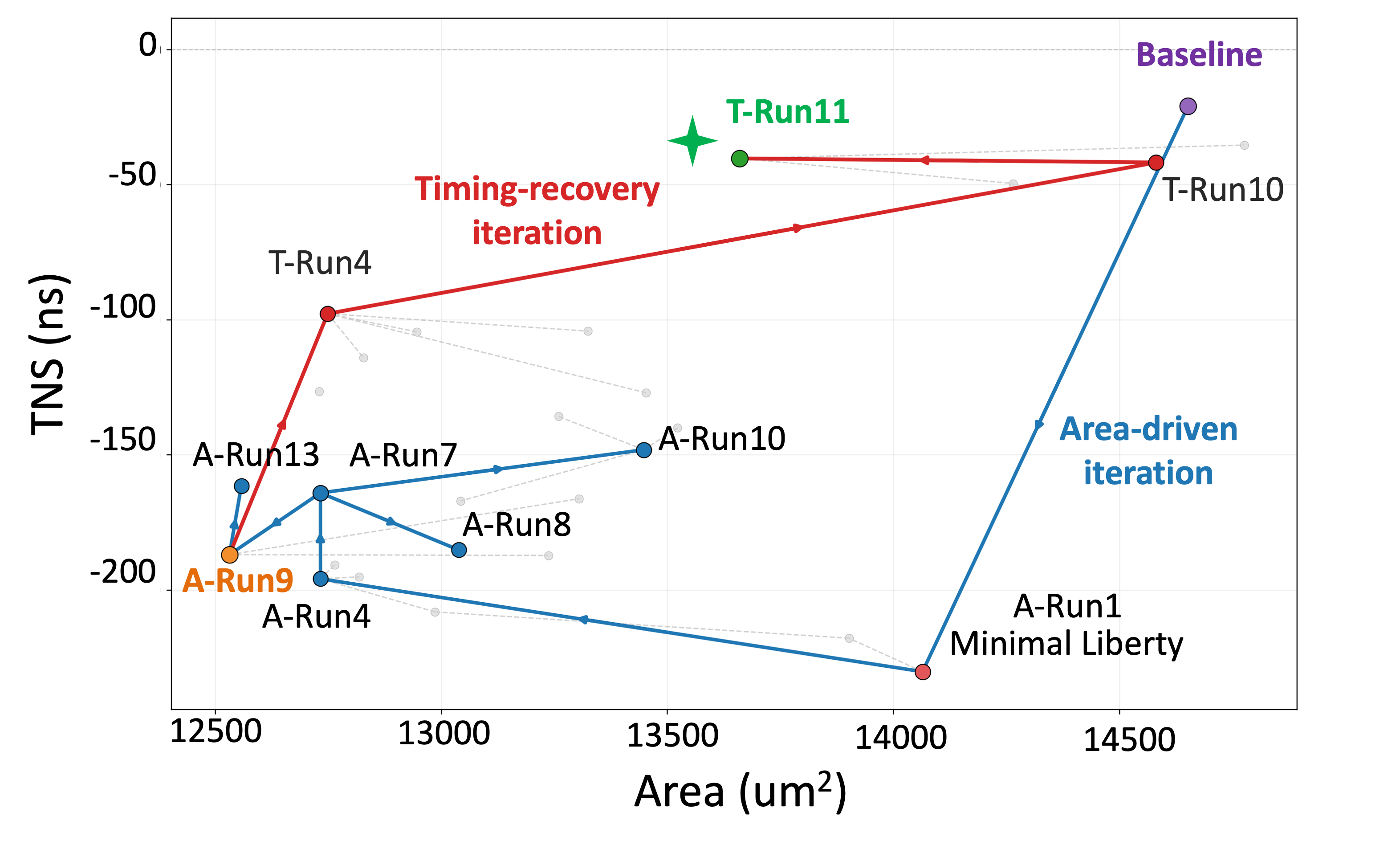}
\caption{Search trajectory for Pareto-driven standard-cell sub-library selection. Blue arrows denote the area-driven exploration stage (\textit{A-Runs}) from the minimal library toward low-area operating points. Red arrows denote the subsequent timing-recovery stage (\textit{T-Runs}), where the agent first performs local probing around \textit{T-Run04} and then transitions to structure-guided restoration, eventually reaching \textit{T-Run11}. Gray dashed arrows indicate alternative local probes.}
\label{fig:lib_subset_search}
\end{figure}

\subsection{Pareto-Driven Standard-Cell Sub-Library Selection}

Our second case study evaluates FluxEDA on standard-cell sub-library specification for early process bring-up. In early technology development, a compact sub-library is often needed to launch the digital flow before the full library is ready~\cite{tsurusaki2025raads}. Such a sub-library reduces layout effort, bring-up cost, and maintenance overhead, but excessive pruning can severely hurt timing and area. The goal is therefore to extract a minimal set of cell families from the full library specification while preserving acceptable implementation quality.

This problem is fundamentally a trade-off among library support, timing, and area. We split the search into two stages: the first stage is area-driven, aiming to identify an area-efficient sub-library by selectively adding the cell support needed for compact mapping (\textit{A-Runs}); the second stage focuses on timing recovery while keeping the library compact (\textit{T-Runs}).

As illustrated in Fig.~\ref{fig:lib_subset_search}, the search begins from the full-library baseline, where the design reaches 14,651.55~$\mu\mathrm{m}^2$ area with WNS/TNS of -0.246/-20.996 ns. Through area-driven exploration, the agent identifies \textit{A-Run9} as a low-area but timing-degraded operating point, reducing area to 12,530.89~$\mu\mathrm{m}^2$ while degrading WNS/TNS to -1.114/-186.934 ns. The agent then branches from this point to perform targeted local recovery, reaching intermediate points such as \textit{T-Run04} by restoring essential inverter-drive flexibility at a modest area cost.

When further one-step local additions show diminishing returns, the agent shifts to deeper critical-path analysis through continuous interaction with the STA tool via FluxEDA. By comparing the critical paths in \textit{T-Run04} with the baseline, the agent diagnoses that the remaining timing loss stems from broader capability limitations of the reduced library, such as insufficient drive strength in shared upstream logic and limited ability to jointly optimize neighboring stages, rather than isolated missing cells. Guided by this diagnosis, the search transitions to structure-guided restoration and eventually reaches \textit{T-Run11}.

Table~\ref{tab:lib_subset_case} summarizes the key milestones and QoR metrics, with the complete iteration trajectory provided in Appendix~\ref{app:lib_subset_traj}. Notably, \textit{T-Run10} already recovers most of the lost timing while keeping the used cell-family set unchanged. The final solution, \textit{T-Run11}, achieves 13,659.95~$\mu\mathrm{m}^2$ area with WNS/TNS of -0.448/-40.306 ns over 157 violating paths, while using only 36 exact cell references across 23 families. Compared with the baseline, this corresponds to 75.8\% and 54.9\% reductions in used references and families, respectively, while still recovering 88.4\% of the TNS gap from \textit{A-Run9}.

Overall, this case study shows that FluxEDA coordinates multiple EDA tools in a stable iterative loop, enabling the agent to move from local probing to structure-guided refinement and converge to a compact yet high-quality sub-library.

\section{Conclusion}
\label{sec:conclu}

We presented FluxEDA, a unified and stateful infrastructure substrate for agentic EDA. Through explicit capability exposure, managed backend instances, and persistent execution context, FluxEDA provides a practical interface between upper-layer agents and heterogeneous EDA environments. Two commercial backend case studies, including post-route timing ECO and standard-cell sub-library optimization, show that the framework can support multi-step iterative optimization over real tool contexts. These results suggest that stateful and governed tool integration is a promising direction for practical agent-assisted EDA automation.


{
  \bibliographystyle{IEEEtran-sim}
  \bibliography{ref/Top-sim,ref/ref}
}
\clearpage
\clearpage
\onecolumn
\appendix
\setcounter{table}{0}   
\renewcommand{\thetable}{A\arabic{table}}
\section{Complete Iteration Trajectory for Sub-Library Selection}
\label{app:lib_subset_traj}

Table~\ref{tab:lib_subset_full_traj} summarizes the complete decision trajectory of the sub-library experiment. \textit{A-Runs} denote the area-driven exploration stage, and \textit{T-Runs} denote the timing-driven recovery stage.

\begin{table*}[htbp]
\centering
\caption{Complete iteration trajectory for Pareto-driven standard-cell sub-library selection.}
\label{tab:lib_subset_full_traj}
\renewcommand{\arraystretch}{1.06}
\setlength{\tabcolsep}{4pt}
\scriptsize
\begin{tabular*}{\textwidth}{@{\extracolsep{\fill}}llp{2.8cm}p{4.2cm}rrrr@{}}
\toprule
ID & Parent & Change & Decision Intent & Area ($\mu\mathrm{m}^2$) & WNS (ns) & TNS (ns) & Viol. Paths \\
\midrule
Baseline & -- & full-library baseline & Reference upper bound with full library support & 14651.546 & -0.246 & -20.996 & 122 \\

A-Run1  & --      & minimal textbook X1 seed & Test whether the smallest valid seed library can launch the flow & 14064.271 & -1.462 & -230.246 & 241 \\
A-Run2  & A-Run1  & add ND3/NR3 X1 & Check whether 3-input logic reduces depth under the minimal seed & 13901.532 & -1.339 & -217.843 & 241 \\
A-Run3  & A-Run2  & add AOI12/OAI12 X1 & Check whether compound gates further compress critical logic depth & 12985.748 & -1.301 & -208.113 & 244 \\
A-Run4  & A-Run3  & add AOI22/OAI22 X1 & Complete a basic compound-gate support set and probe the area minimum & 12732.037 & -1.145 & -195.882 & 244 \\
A-Run5  & A-Run4  & add OAI12 X2 & Test whether endpoint-side complex-gate resizing helps & 12763.372 & -1.179 & -190.739 & 244 \\
A-Run6  & A-Run4  & add ND3 X2 & Test whether stronger 3-input drive improves timing efficiently & 12817.955 & -1.108 & -195.168 & 244 \\
A-Run7  & A-Run4  & add NR2 X2 & Probe whether high-frequency NR2 is a high-information gain lever & 12732.037 & -0.938 & -164.075 & 240 \\
A-Run8  & A-Run7  & add AOI12 X2 & Check whether further AOI12 support is worthwhile on the A-Run7 branch & 13038.309 & -1.096 & -185.193 & 238 \\
A-Run9  & A-Run7  & add OAI22 X2 & Establish the leftmost low-area Pareto operating point & 12530.888 & -1.114 & -186.934 & 237 \\
A-Run10 & A-Run7  & add ND2 X2 & Push the frontier toward a more timing-oriented alternative & 13447.683 & -0.842 & -148.127 & 242 \\
A-Run11 & A-Run9  & add ND2 X2 & Test whether ND2 support helps on the low-area branch & 13236.426 & -1.137 & -187.231 & 240 \\
A-Run12 & A-Run9  & add OAI12 X2 & Test whether endpoint-side complex-gate support helps on the low-area branch & 13304.150 & -0.932 & -166.229 & 237 \\
A-Run13 & A-Run9  & add NR2 X3 & Refine NR2 drive on the low-area branch & 12557.168 & -1.031 & -161.515 & 233 \\

T-Run01 & A-Run10 & add NR2X3 & Test whether the NR2 backbone of A-Run10 can push timing further & 13522.482 & -0.830 & -139.982 & 240 \\
T-Run02 & A-Run10 & add OAI12X2 & Test whether stronger endpoint-side compound gates help on A-Run10 & 13042.352 & -0.990 & -167.045 & 242 \\
T-Run03 & A-Run10 & add ND2X3 & Test whether ND2 strength is still insufficient on A-Run10 & 13258.664 & -0.784 & -135.774 & 238 \\
T-Run04 & A-Run9  & add INVX2 & Test whether the low-area branch mainly lacks basic inverter-drive flexibility & 12748.210 & -0.791 & -97.780 & 228 \\
T-Run05 & T-Run04 & add INVX2, NR2X3 & Continue pushing NR2 on top of T-Run04 & 12945.316 & -0.764 & -104.496 & 225 \\
T-Run06 & T-Run04 & add INVX2, OAI12X2 & Add endpoint-side compound-gate support on top of T-Run04 & 13324.366 & -0.697 & -104.212 & 236 \\
T-Run07 & T-Run03 & add ND2X3, BUFX2 & Test whether shared buffer/drive support yields a lower-area timing compromise & 12729.004 & -0.904 & -126.508 & 239 \\
T-Run08 & T-Run04 & add INVX2, BUFX2 & Test whether a shared electrical segment is the dominant TNS bottleneck & 12827.052 & -0.807 & -114.012 & 236 \\
T-Run09 & T-Run04 & add INVX2, ND2X2 & Test whether high-frequency ND2 can further reduce globals TNS & 13452.737 & -0.856 & -127.045 & 229 \\
T-Run10 & T-Run04 & restore shared upstream-drive pack & Recover the stronger shared upstream-drive capability indicated by path analysis & 14580.790 & -0.385 & -41.803 & 181 \\
T-Run11 & T-Run04 & shared-drive pack + endpoint-tail pack & Restore upstream shared drive together with endpoint-critical logic for balanced recovery & 13659.951 & -0.448 & -40.306 & 157 \\
T-Run12 & T-Run11 & remove OAI12X6, AOI13X3 & Test whether low-use tail cells are dispensable & 14775.874 & -0.322 & -35.381 & 172 \\
T-Run13 & T-Run11 & remove OAI12X6, AOI13X3, AOI22X2 & Further trim the T-Run11 tail support for lower area & 14264.410 & -0.363 & -49.588 & 228 \\
\bottomrule
\end{tabular*}
\end{table*}
\end{document}